\documentclass[12pt,journal,draftclsnofoot,a4paper,oneside,onecolumn]{IEEEtran}

\usepackage{amssymb}
\usepackage{amsmath}
\usepackage{graphicx}
\usepackage{cite}
\usepackage{url}

\begin{document}

\title{Physical Layer Security for Two-Way Untrusted Relaying with Friendly Jammers}

\author{
\IEEEauthorblockN{Rongqing Zhang\IEEEauthorrefmark{1}, Lingyang
Song\IEEEauthorrefmark{1}, Zhu Han\IEEEauthorrefmark{2}, and Bingli
Jiao\IEEEauthorrefmark{1}\\}
\IEEEauthorblockA{\IEEEauthorrefmark{1}\normalsize{School of Electrical Engineering and Computer Science, Peking University, Beijing, China.}\\
                  \IEEEauthorrefmark{2}\normalsize{Electrical and Computer Engineering Department, University of Houston, Houston, USA.}}
\thanks{This work is partially supported by US NSF CNS-0910401,
CNS-0905556, and CNS-0953377.} }

\maketitle

\begin{abstract}

In this paper, we consider a two-way relay network where two sources
can communicate only through an untrusted intermediate relay, and
investigate the physical layer security issue of this two-way relay
scenario. Specifically, we treat the intermediate relay as an
eavesdropper from which the information transmitted by the sources
needs to be kept secret, despite the fact that its cooperation in
relaying this information is essential. We indicate that a non-zero
secrecy rate is indeed achievable in this two-way relay network even
without external friendly jammers. As for the system with friendly
jammers, after further analysis, we can obtain that the secrecy rate
of the sources can be effectively improved by utilizing proper
jamming power from the friendly jammers. Then, we formulate a
Stackelberg game model between the sources and the friendly jammers
as a power control scheme to achieve the optimized secrecy rate of
the sources, in which the sources are treated as the sole buyer and
the friendly jammers are the sellers. In addition, the optimal
solutions of the jamming power and the asking prices are given and a
distributed updating algorithm to obtain the Stakelberg equilibrium
is provided for the proposed game. Finally, the simulations results
verify the properties and the efficiency of the proposed Stackelberg
game based scheme.

\end{abstract}

\IEEEpeerreviewmaketitle

\newpage

\section{Introduction}%

Traditionally security in wireless networks has been mainly
considered at higher layers using cryptographic methods. However,
recent advances in wireless decentralized and ad-hoc networking have
led to an increasing attention on studying physical layer based
security. The basic idea of physical layer security is to exploit
the physical characteristics of the wireless channel to provide
secure communication. The security is quantified by the
\emph{secrecy capacity}, which is defined as the maximum rate of
reliable information sent from the source to the intended
destination in the presence of eavesdroppers. This line of work was
pioneered by Aaron Wyner, who introduced the wiretap channel and
established fundamental results of creating perfectly secure
communications without relying on private keys \cite{Wyner-1975}.
Wyner showed that when the eavesdropper channel is a degraded
version of the main channel, the source and the destination can
exchange perfectly secure messages at a non-zero rate. In follow-up
work \cite{LH-1978}, the secrecy capacity of Gaussian wiretap
channel was studied, and in \cite{CK-1978} Wyner's approach was
extended to the transmission of confidential messages over broadcast
channels. Recently, researches on physical layer security have
generalized these studies to wireless fading channels
\cite{PB-2005,BR-2006,LPS-2008,GLG-2008}, MIMO channels
\cite{NG-2005,SU-2007,KWWE-2007,OH-2008,SLU-2009}, and various
multiple access scenarios
\cite{LP-2006,LP-2008,CN-2008,KTW-2008,TY-2008,LLPS-2009}.

Motivated the fact that if the source-wiretapper channel is less
noisy than the source-destination channel, the perfect secrecy
capacity will be zero \cite{CK-1978}, some recent work has been
proposed to overcome this limitation using relay cooperation, which
mainly consists of \emph{cooperative relaying}
\cite{DHPP-2008,DHPP-2009}, and \emph{cooperative jamming}
\cite{DHPP-2010,LG-2008}. For instance, in \cite{DHPP-2008} and
\cite{DHPP-2009}, the authors proposed effective decode-and-forward
(DF) and amplify-and-forward (AF) based cooperative relaying
protocols for physical layer security, respectively. Cooperative
jamming is another approach to improve the secrecy capacity by
distracting the eavesdropper with codewords independent of the
source messages. In \cite{DHPP-2010} and \cite{LG-2008}, several
cooperative jamming schemes were investigated for different
scenarios, where classical relay strategies fail to offer positive
performance gains. Relay channel with confidential messages was
studied in \cite{Oohama-2001} and \cite{Oohama-2007}, where the
relay node acts both as an eavesdropper and a helper. In
\cite{HY-2010}, it was established that cooperation even with an
untrusted relay node could be beneficial in relay channels with
orthogonal components. Then in \cite{HY-2008}, the authors
considered a two-hop communication system using an untrusted relay
and showed that a cooperative jammer enables a positive secrecy
capacity which would be otherwise impossible.

Two-way communication is a common scenario where two terminals
transmit information to each other simultaneously. Recently, the
two-way relay channel
\cite{RW-2006,RW-2007,CHK-2009,ZLCC-2009,HH-2006} has attracted lots
of interest from both academic and industrial communities due to its
advantage in saving bandwidth efficiently. In \cite{RW-2006} and
\cite{RW-2007}, both AF and DF protocols for one-way relay channels
were extended to general full-duplex discrete two-way relay channel
and half-duplex Gaussian two-way relay channel, respectively. In
\cite{CHK-2009}, different relay strategies consisting of AF, DF,
and EF (estimate-and-forword) for uncoded two-way relay channels
were investigated. In \cite{ZLCC-2009}, analogue network coding
based two-way relay channel with linear processing was analyzed and
an optimal relay beamforming structure was presented. In
\cite{HH-2006}, a joint network-channel coding was proposed for the
two-way relay channel, where channel codes are used at both the
sources and a network code is used at the intermediate relay.
Although two-way relay networks have received so much attention so
far, the security issue about the relay, especially from the
physical layer security point of view, has not been well
investigated.

To improve the security in two-way relay channel, distributed
protocols are desired. Game theory \cite{FT-GameTheory-1993} offers
a formal analytical framework with a set of mathematical tools to
study the complex interactions among interdependent rational
players. Recently, there has been significant growth in research
activities that use game theory for analyzing communication
networks, mainly due to the need for developing autonomous,
distributed, and flexible mobile networks where the network devices
can make independent and rational strategic decisions, as well as
the need for low complexity distributed algorithms for competitive
or collaborative scenarios \cite{Han-GameTheory-2010}. In
\cite{HPMWWB-2008} and \cite{JLLP-2009}, the authors introduced some
recent studies on signal processing and communication networks using
game theory. In \cite{HMDH-Journal-2009} and \cite{HMDH-2009}, the
authors employed game theory to physical layer security to study the
interaction between the source and the jammers who assist the source
by distracting the eavesdropper, and got some distributed game
solutions. For two-way relay networks, it is desirable to study the
physical layer security problems with the aid of game theory similar
to those for one-way relay cases.

In this paper, we investigate physical layer security issues in a
two-way relay network with friendly jammers. The two sources can
exchange information only through an untrusted relay, as there is no
direct communication link between them. The untrusted AF relay acts
as both an essential relay and a malicious eavesdropper that has the
incentive to eavesdrop on the information transmission. For
convenience and ease of comparison, we first study the system
without friendly jammers as a special case. We find that a non-zero
secrecy rate here is indeed available even without the help of
friendly jammers. We also derive an optimal power vector of the
relay and the sources by maximizing the secrecy rate. We further
investigate the two-way relay secure communication with friendly
jammers, and find that a positive gain can be obtained in the
secrecy rate by utilizing proper jamming power from the friendly
jammers. Then the problem comes to how to effectively utilize the
jamming power from different friendly jammers to maximize the
secrecy rate. Thus, we propose a Stackelberg game model between the
sources and the friendly jammers as a power control scheme. In the
defined game, the sources pay the friendly jammers for interfering
the untrusted relay in order to increase the secrecy rate, while the
friendly jammers charge the sources with a certain price for their
service of jamming. In addition, the optimal solutions of the
jamming power and the asking prices are given and a distributed
updating algorithm to obtain the Stakelberg equilibrium is provided
for the proposed game. Furthermore, a centralized scheme is also
proposed for comparison with the distributed Stackelberg game based
scheme. Finally, the proposed approaches and solutions are verified
by simulations.

The rest of this paper is organized as follows. In Section
\uppercase\expandafter{\romannumeral2}, the system model of two-way
relay communication with friendly jammers is described and the
corresponding secrecy rate is formulated. In Section
\uppercase\expandafter{\romannumeral3}, a two-way relay system
without jammers as a special case is investigated. In Section
\uppercase\expandafter{\romannumeral4}, we define a Stackelberg type
of buyer/seller game to investigate the interaction between the
sources and the friendly jammers, and analyze the optimization
problem of physical layer security in the presence of friendly
jammers. Simulation results are provided in Section
\uppercase\expandafter{\romannumeral5}, and the conclusions are
drawn in Section \uppercase\expandafter{\romannumeral6}.

\section{System Model}%

\begin{figure}[!t]
\centering
\includegraphics[width=4.8in]{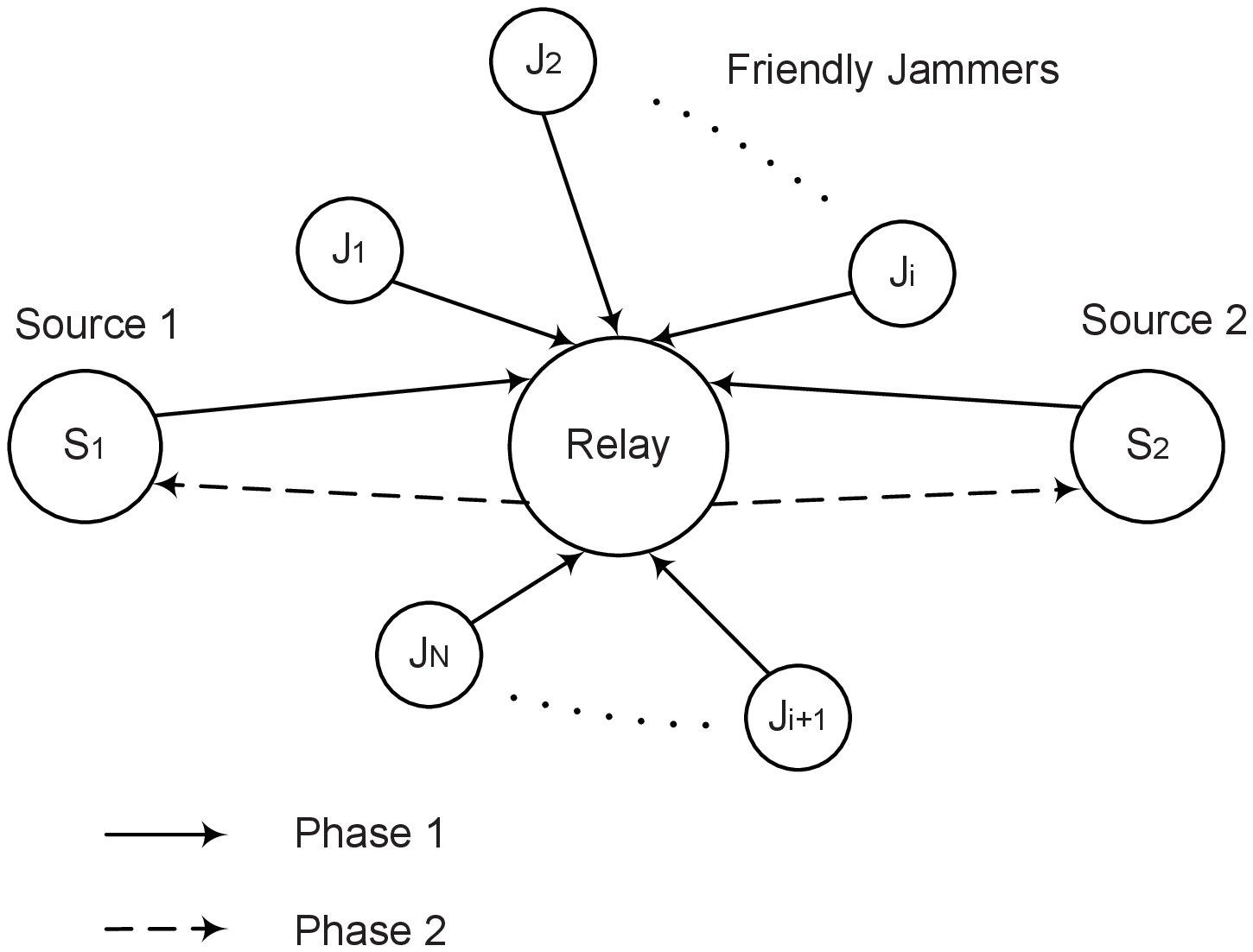}
\caption{System model for two-way relay communications with friendly
jammers} \label{fig_1}
\end{figure}

As shown in Fig. \ref{fig_1}, we consider a basic two-way relay
network consisting of two source nodes, one untrusted relay node,
and $N$ friendly jammer nodes, which are denoted by $S_k$, $k=1,2$,
$R$, and $J_i$, $i=1,2,\ldots,N$, respectively. We denote by
$\mathcal{N}$ the set of indices $\{1,2,\ldots,N\}$. All the nodes
here are equipped with only a single omni-directional antenna and
operate in a half-duplex way, i.e., each node cannot receive and
transmit simultaneously. Then the complete transmission can be
divided into two phases. During the first phase, shown with solid
lines, both source nodes transmit their information to the relay
node. Simultaneously, the friendly jammers also transmit the jamming
signals in order to distract the malicious relay. In the second
phase, shown with dashed lines, the relay node broadcasts a combined
version of the received signals to both source nodes. Note that in
the system we investigate there is only one intermediate relay and
we assume that no direct link exists between the two sources. Thus,
the sole untrusted relay is necessary for the two-way relaying data
transmission. A key assumption \footnote{We can guarantee this by
using some pseudo-random codes which are known to both the friendly
jammers and the sources but not open to the malicious relay. Beyond
this, we can also use some cryptographic signals at the friendly
jammers for jamming, where the decryption book is a secret key only
open to the sources. Then the sources can have a perfect knowledge
of the jamming signals if each jammer sends some additional bits
consisting of the information of the jamming signal transmitted
(e.g., which code or which encryption method to be used). The
information that needs to be sent is for one time, which will lead
to trivial bandwidth cost.} we make here is that the sources have
perfect knowledge of the jamming signals transmitted by the friendly
jammers, for they have paid for the service.

Let $s_1$, $s_2$, and $s_i^J$, $i\in\mathcal{N}$, denote the signal
to be transmitted by the source $S_1, S_2$, and the jammers $J_i$,
$i\in\mathcal{N}$, respectively. Suppose source nodes $S_1$ and
$S_2$ transmit with power $p_1$ and $p_2$, and the channel gains
from the source nodes to the relay node are denoted by $h_{S_k,R}$,
$k=1,2$. Each friendly jammer node $i$ transmits with power $p_i^J$,
and the channel gain from it to the relay node is denoted by
$h_{J_i,R}$, $i\in\mathcal{N}$. The channel gain contains the path
loss and the Rayleigh fading coefficient with zero mean and unit
variance. For simplicity, we assume that the fading coefficients are
constant over one frame, and vary independently from one frame to
another.

In phase $1$, the received signal at the malicious relay can be
expressed as
\begin{align} \label{eq1}
{y_r} = n_r+\sqrt{p_1} s_1 h_{S_1,R}+\sqrt{p_2} s_2 h_{S_2,R}+\sum
_i \sqrt{p_i^J} s_i^J h_{J_i,R},
\end{align}
where $n_r$ denotes the thermal noise at the relay node $R$, which
is a zero mean Gaussian random variable with two sided power
spectral density of $\sigma^2$, i.e.,
$\mathcal{C}\mathcal{N}(0,\sigma^2)$. Furthermore, we assume that
the noises at $S_1$, $S_2$, and $R$ are independent and identically
distributed (i.i.d.).

In phase $2$, the malicious relay, which works in AF mode, amplifies
the received signal $y_r$ by a factor $\beta$ and then broadcasts
the signal to both $S_1$ and $S_2$ with power $p_r$. The power
normalization factor $\beta$ at the relay node can be written as
\begin{align} \label{eq2}
{\beta} =
\left(p_1\left|h_{S_1,R}\right|{}^2+p_2\left|h_{S_2,R}\right|{}^2+\sum
_i p_i^J\left|h_{J_i,R}\right|{}^2+\sigma ^2\right)^{-1/2}.
\end{align}
Then the corresponding signal received by the source $S_1$ , denoted
by $y_1$, can be written as
\begin{align} \label{eq3}
{y_1} &= \beta \sqrt{p_r}h_{S_1,R}y_r + n_1 =\xi _1s_1+\upsilon
_1s_2+\sum _i \mu _{1,i}s_i^J+\omega _1,
\end{align}
where ${\xi _1} \triangleq \beta \sqrt {{p_r}{p_1}} h_{{S_1},R}^2$,
${\upsilon _1} \triangleq \beta \sqrt {{p_r}{p_2}}
{h_{{S_1},R}}{h_{{S_2},R}}$, ${\mu _{1,i}} \triangleq \beta \sqrt
{{p_r}p_i^J} {h_{{J_i},R}}{h_{{S_1},R}}$, and ${\omega _1}
\triangleq \beta \sqrt {{p_r}} {h_{{S_1},R}}{n_r} + {n_1}$.
Similarly, the signal received by the source $S_2$, denoted by
$y_2$, can be written as
\begin{align} \label{eq4}
{y_2} &= \beta \sqrt {{p_r}} {h_{{S_2},R}}{y_r} + {n_2} = {\xi
_2}{s_1} + {\upsilon _2}{s_2} + \sum\limits_i {{\mu _{2,i}}s_i^J}  +
{\omega _2},
\end{align}
where ${\xi _2} \triangleq \beta \sqrt {{p_r}{p_1}}
{h_{{S_1},R}}{h_{{S_2},R}}$, ${\upsilon _2} \triangleq \beta \sqrt
{{p_r}{p_2}} h_{{S_2},R}^2$, ${\mu _{2,i}} \triangleq \beta \sqrt
{{p_r}p_i^J} {h_{{J_i},R}}{h_{{S_2},R}}$, and ${\omega _2}
\triangleq \beta \sqrt {{p_r}} {h_{{S_2},R}}{n_r} + {n_2}$.

Assuming that both the source nodes and the jammer nodes are
independent, from (\ref{eq1}), in phase $1$, using the matched
filter (MF) \footnote{For simplicity, we use the matched filter for
signal detection \cite{Verdu-MultiuserDetection-1998} while many
other advanced detectors can be applied and the analysis can be done
in a similar way.}, the untrusted relay node has the capacity with
respect to $S_1$ and $S_2$ as
\begin{align} \label{eq5}
{C_1^m} = \frac{W}{2}\log \left( {1 +
\frac{{{p_1}{g_{{S_1},R}}}}{{{\sigma ^2} + {p_2}{g_{{S_2},R}} +
\sum\limits_i {p_i^J{g_{{J_i},R}}} }}} \right),
\end{align}
and
\begin{align} \label{eq6}
{C_2^m} = \frac{W}{2}\log \left( {1 +
          \frac{{{p_2}{g_{{S_2},R}}}}{{{\sigma ^2} + {p_1}{g_{{S_1},R}} +
          \sum\limits_i {p_i^J{g_{{J_i},R}}} }}} \right),
\end{align}
where $W$ represents the channel bandwidth, ${g_{{S_1},R}}
\triangleq {\left| {{h_{{S_1},R}}} \right|^2}$, ${g_{{S_2},R}}
\triangleq {\left| {{h_{{S_2},R}}} \right|^2}$, and ${g_{{J_i},R}}
\triangleq {\left| {{h_{{J_i},R}}} \right|^2}$, $i\in\mathcal{N}$.

In phase $2$, at $S_1$, as $s_1$ as well as $s_i^J$ is known to the
source node, and thus we have
\begin{align} \label{eq7}
\tilde{y}_1 = {\upsilon _1}{s_2} + {\omega _1}.
\end{align}
Then, the corresponding SNR for the transmission from $S_2$ to
$S_1$, denoted by $\gamma_2$, can be expressed as
\begin{align} \label{eq8}
{\gamma _2} &= \frac{{{{\left| {{\upsilon _1}}
\right|}^2}}}{{Var\left\{ {{\omega _1}} \right\}}} =
\frac{{{p_2}{g_{{S_2},R}}}}{{{\sigma ^2} + {K_2} + \sum\limits_i
{\frac{{{\sigma ^2}{g_{{J_i},R}}}}{{{p_r}{g_{{S_1},R}}}}p_i^J} }},
\end{align}
where ${K_2} = \frac{{{\sigma ^2}({p_1}{g_{{S_1},R}} +
{p_2}{g_{{S_2},R}} + {\sigma ^2})}}{{{p_r}{g_{{S_1},R}}}}$.
Similarly, at $S_2$, the received signal with $s_2$ and $s_i^J$
removed can be written as
\begin{align} \label{eq9}
\tilde{y}_2 = {\xi _2}{s_1} + {\omega _2}.
\end{align}
The corresponding SNR for the transmission from $S_1$ to $S_2$,
denoted by $\gamma_1$, can be expressed as
\begin{align} \label{eq10}
{\gamma _1} &= \frac{{{{\left| {{\xi _2}} \right|}^2}}}{{Var\left\{
{{\omega _2}} \right\}}} = \frac{{{p_1}{g_{{S_1},R}}}}{{{\sigma ^2}
+ {K_1} + \sum\limits_i {\frac{{{\sigma
^2}{g_{{J_i},R}}}}{{{p_r}{g_{{S_2},R}}}}p_i^J} }},
\end{align}
where ${K_1} = \frac{{{\sigma ^2}\left( {{p_1}{g_{{S_1},R}} +
{p_2}{g_{{S_2},R}} + {\sigma ^2}} \right)}}{{{p_r}{g_{{S_2},R}}}}$.

Capacities of two-way relay channel between the two sources are
denoted by $C_1$ and $C_2$, and we have
\begin{align} \label{eq11}
{C_1} = \frac{W}{2}\log \left( {1 + {\gamma _1}} \right),
\end{align}
and
\begin{align} \label{eq12}
{C_2} = \frac{W}{2}\log \left( {1 + {\gamma _2}} \right).
\end{align}
Then, the secrecy rate for $S_1$ and $S_2$ \cite{BR-2006} can be
defined as
\begin{align} \label{eq13}
{C_1^s} = {\left( {{C_1} - C_1^m} \right)^ + },
\end{align}
and
\begin{align} \label{eq14}
{C_2^s} = {\left( {{C_2} - C_2^m} \right)^ + },
\end{align}
where ${\left( x \right)^ + }$ represents $\max \left\{ {x,0}
\right\}$. According to \cite{HY-2010} and \cite{HY-2008}, we have
that the defined secrecy rate is achievable in a two-hop secure
communication with an untrusted relay.

As a special case, if the jammers are not used, the jammers'
transmit power $p_i^J$ should be set to zero, $\forall
i\in\mathcal{N}$. Then from the derivation above, we can get the
corresponding secrecy rate in this case as
\begin{align} \label{eq15}
{\tilde C_1^s} = \frac{W}{2}{\left[ {\log \left( {1 +
\frac{{{p_1}{g_{{S_1},R}}}}{{{\sigma ^2} + {K_1}}}} \right) - \log
\left( {1 + \frac{{{p_1}{g_{{S_1},R}}}}{{{\sigma ^2} +
{p_2}{g_{{S_2},R}}}}} \right)} \right]^ + },
\end{align}
and
\begin{align} \label{eq16}
{\tilde C_2^s} = \frac{W}{2}{\left[ {\log \left( {1 +
\frac{{{p_2}{g_{{S_2},R}}}}{{{\sigma ^2} + {K_2}}}} \right) - \log
\left( {1 + \frac{{{p_2}{g_{{S_2},R}}}}{{{\sigma ^2} +
{p_1}{g_{{S_1},R}}}}} \right)} \right]^ + }.
\end{align}

In the system we investigate, there is only one intermediate relay,
thus this sole relay is necessary in our assumption for two-way
relaying data transmission. Actually, the untrusted relay has the
incentive to forward the signals from both the sources since it can
eavesdrop on the information transmission through this kind of
cooperation. If the relay is non-cooperative that it only receives
but not relays the information, then the problem comes to
deny-of-service attack. However, this can be easily detected by the
sources, then the non-cooperative relay will be treated as a
thorough eavesdropper and lose the good opportunity to eavesdrop on
the information transmission. The sources will then turn to another
intermediate relay for help to relay their information in a
practical scenario where there exist multiple intermediate relays.
In this paper, we focus on the studies how to prevent the untrusted
but necessary intermediate relay from eavesdropping the information,
and thus, for simplicity and without loss of generality, we assume
that there is only one necessary intermediate relay in the system
and the relay is cooperative.

\section{Secrecy Rate of Two-Way Relay Channel Without Jammers}%

For comparison and consistence, we first investigate the special
case without the presence of jammers in this section. We prove that
there indeed exists a positive secrecy rate for the two-way relay
channel even without the help of friendly jammers distracting the
malicious relay. Furthermore, we also obtain an optimal power
allocation of the sources and the relay to maximize the secrecy
rate. In the next section, we will compare the case with friendly
jammers with this case to expect a positive performance gain in the
secrecy rate.

\subsection{Existence of Non-zero Secrecy Rate}

When the eavesdropper channels from the two sources to the malicious
relay are degraded versions of the equivalent main two-way relay
channel between $S_1$ and $S_2$, the two sources can exchange
perfectly secure messages at a non-zero rate. Firstly, we consider
the transmission from $S_1$ to $S_2$. In phase $1$, the malicious
relay receives the signal $s_1$ from $S_1$, which consists of the
information for $S_2$. Meanwhile, $S_2$ also transmits the signal
$s_2$ at the relay, which acts as both the information carrier for
$S_1$ and a jamming signal that makes the eavesdropper channel from
$S_1$ to the malicious relay getting worse. In phase $2$, the
combined signal consisting of $s_1$ and $s_2$ arrives at $S_2$. As
$S_2$ has a perfect knowledge of its own signal $s_2$, the signal
that jammed the malicious relay in phase $1$ has no such an effect
on $S_2$. Therefore, it makes possible that the eavesdropper channel
is worse than the data transmission channel from $S_1$ to $S_2$,
which means that a non-zero rate for secure communication from $S_1$
to $ S_2$ is available. It is the same situation in the transmission
from $S_2$ to $S_1$. From (\ref{eq15}), (\ref{eq16}) and the
expressions of $K_1$ in (\ref{eq10}) and $K_2$ in (\ref{eq8}), we
can write the probability of the existence of a non-zero secrecy
rate as
\begin{align} \label{eq17}
{\rm P}\left( {\tilde C_1^s > 0,\tilde C_2^s > 0} \right) &= {\rm P}\left( {{K_1} < {p_2}{g_{{S_2},R}},{K_2} < {p_1}{g_{{S_1},R}}} \right) \nonumber \\
&= {\rm P}\left( {{p_r} > \max \left\{
{\frac{T}{{{p_2}g_{{S_2},R}^2}},\frac{T}{{{p_1}g_{{S_1},R}^2}}}
\right\}} \right),
\end{align}
where $T = ({p_1}{g_{{S_1},R}} + {p_2}{g_{{S_2},R}} + {\sigma
^2}){\sigma ^2}$.

Considering the power constraints ${p_1} \le {p_{max}}$, ${p_2} \le
{p_{max}}$, and ${p_r} \le {p_{max}}$, we can get that there exists
at least one pair of $\left( {{p_r},{p_1},{p_2}} \right)$ that
satisfies ${p_r} > \max \left\{
{\frac{T}{{{p_2}g_{{S_2},R}^2}},\frac{T}{{{p_1}g_{{S_1},R}^2}}}
\right\}$, under the channel condition of $\frac{{{g_{{S_1},R}} +
{g_{{S_2},R}}}}{{{g_{{S_1},R}}{g_{{S_2},R}}}} <
\frac{{{p_{max}}}}{{{\sigma ^2}}}$. Therefore, we have ${\rm
P}\left( {\tilde C_1^s > 0,\tilde C_2^s > 0} \right) > 0$ at some
power vectors of $\left( {{p_r},{p_1},{p_2}} \right) $, which
actually indicates that a non-zero secrecy rate in the two-way relay
channel is indeed available.

\subsection{Maximizing the Secrecy Rate}

In this subsection, we try to get an optimal power vector of $\left(
{{p_r},{p_1},{p_2}} \right)$ which maximizes the secrecy rate of the
two-way relay channel. We can formulate the problem subject to the
individual secrecy rate constraints and power constraints as
\begin{align} \label{eq18}
&\max \;{\tilde C^s} = \max \;\sum\limits_{k = 1}^2 {\tilde C_k^s}, \\
\mbox{s.t.}\quad &\left\{ \begin{array}{l}
\tilde C_1^s > 0,\:\tilde C_2^s > 0 \\
{p_1} \le {p_{max}},\:{p_2} \le {p_{max}},\:{p_r} \le {p_{max}} \\
\end{array} \right.. \nonumber
\end{align}

From (\ref{eq17}), we have that
\begin{align} \label{eq19}
\tilde C_1^s > 0,\;\tilde C_2^s > 0 \Leftrightarrow {p_r} > \max
\left\{
{\frac{T}{{{p_2}g_{{S_2},R}^2}},\frac{T}{{{p_1}g_{{S_1},R}^2}}}
\right\}.
\end{align}
From (\ref{eq15}), (\ref{eq16}), and (\ref{eq18}), we can get that
\begin{align} \label{eq20}
{\tilde C^s} = \frac{W}{2}{\left( {\log \tilde F\left(
{{p_r},{p_1},{p_2}} \right)} \right)^ + },
\end{align}
where
\begin{align} \label{eq21}
\tilde F\left( {{p_r},{p_1},{p_2}} \right) \triangleq \frac{{\left(
{1 + \frac{{{p_1}{g_{{S_1},R}}}}{{{\sigma ^2} + {K_1}}}}
\right)\left( {1 + \frac{{{p_2}{g_{{S_2},R}}}}{{{\sigma ^2} +
{K_2}}}} \right)}}{{\left( {1 + \frac{{{p_1}{g_{{S_1},R}}}}{{{\sigma
^2} + {p_2}{g_{{S_2},R}}}}} \right)\left( {1 +
\frac{{{p_2}{g_{{S_2},R}}}}{{{\sigma ^2} + {p_1}{g_{{S_1},R}}}}}
\right)}}.
\end{align}

As $\tilde F({p_r},{p_1},{p_2})$ has the same monotonic property as
${\tilde C^s}$ under the conditions of (\ref{eq18}), we can
transform the optimization problem as
\begin{align} \label{eq22}
&\max \;\tilde F\left( {{p_r},{p_1},{p_2}} \right), \\
\mbox{s.t.}\quad &\left\{ \begin{array}{l}
{p_r} > \max \left\{ {\frac{T}{{{p_2}g_{{S_2},R}^2}},\frac{T}{{{p_1}g_{{S_1},R}^2}}} \right\} \\
{p_1} \le {p_{max}},\:{p_2} \le {p_{max}},\:{p_r} \le {p_{max}} \\
\end{array} \right.. \nonumber
\end{align}

It can be calculated that $\frac{{\partial \tilde F\left(
{{p_r},{p_1},{p_2}} \right)}}{{\partial {p_r}}} > 0$ is always
established under the conditions of (\ref{eq22}), which implies that
$\tilde F\left( {{p_r},{p_1},{p_2}} \right)$ is a monotonically
increasing function of ${p_r}$. Therefore, when maximizing the
secrecy rate ${\tilde C^s}$, ${p_{r\_opt}} = {p_{max}}$, where
${p_{r\_opt}}$ denotes the optimal relay power \footnote{Note that
here we calculate the optimal power solution of $p_r$ only from a
mathematical perspective to maximize the secrecy rate. In fact, the
intermediate relay has no incentive to transmit with the maximum
power.}. As a result, the problem can be further transformed into
$\max \;\tilde F\left( {{p_{max}},{p_1},{p_2}} \right)$.

The optimal solutions of $p_1$ and $p_2$ when maximizing the secrecy
rate can be easily obtained under different conditions (i.e.,
$g_{S_1,R} > g_{S_2,R}$, $g_{S_1,R} < g_{S_2,R}$, and $g_{S_1,R} =
g_{S_2,R}$) through the Lagrangian method by solving the
Karush-Kuhn-Tucker (KKT) conditions \cite{BV-Convex-2006}. In this
paper, subject to the space limit, we omit the detailed computing
process and only give the results of the optimal solutions of $p_1$
and $p_2$ as:


\begin{enumerate}

\item
For the case that ${g_{{S_1},R}} > {g_{{S_2},R}}$, it yields that
${p_{2\_opt}} = {p_{max}}$. Meanwhile, if there exists a solution
$p_1^* \in \left( {0,{p_{max}}} \right]$ that satisfies the equation
$\frac{{\partial \tilde F\left( {{p_{max}},{p_1},{p_{max}}}
\right)}}{{\partial {p_1}}} = 0$, then we have ${p_{1\_opt}} =
p_1^*$. Otherwise, we have ${p_{1\_opt}} = {p_{max}}$. Here
${p_{1\_opt}}$ and ${p_{2\_opt}}$ denote the optimal power
transmitted by ${S_1}$ and ${S_2}$, respectively.

\item
For the case that ${g_{{S_1},R}} < {g_{{S_2},R}}$, it yields that
${p_{1\_opt}} = {p_{max}}$. Meanwhile, if there exists a solution
$p_2^* \in \left( {0,{p_{max}}} \right]$ that satisfies the equation
$\frac{{\partial \tilde F\left( {{p_{max}},{p_{max}},{p_2}}
\right)}}{{\partial {p_2}}} = 0$,  then we have ${p_{2\_opt}} =
p_2^*$. Otherwise, we have ${p_{2\_opt}} = {p_{max}}$.

\item
For the case that ${g_{{S_1},R}} = {g_{{S_2},R}}$, we have that
${p_{1\_opt}} = {p_{max}}$, and ${p_{2\_opt}} = {p_{max}}$.

\end{enumerate}

\begin{figure}[ht]
\centering
\includegraphics[width=4.8in]{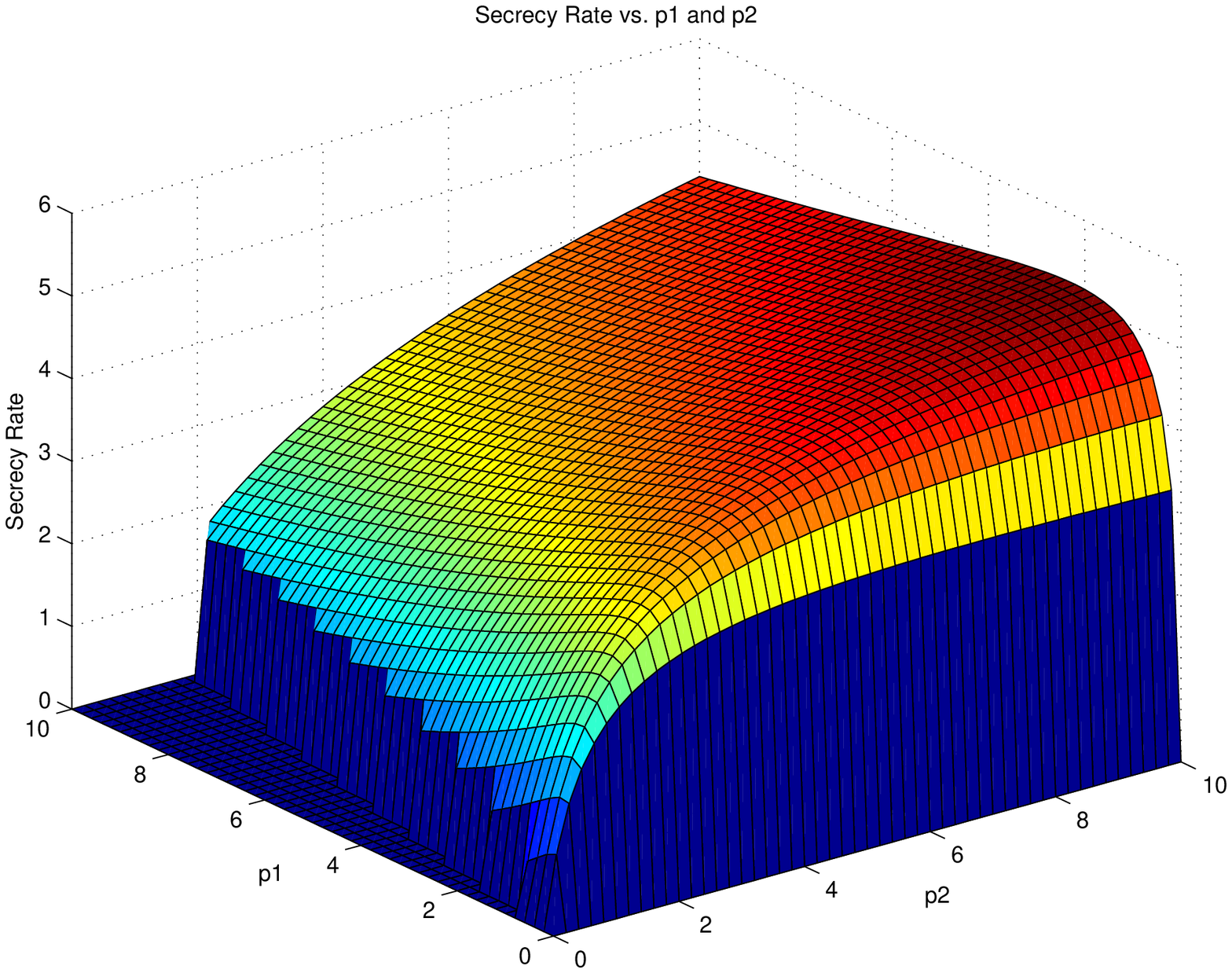}
\caption{Secrecy rate vs. $p_1$ and $p_2$ for the case without
jammers when ${g_{{S_1},R}} > {g_{{S_2},R}}$} \label{fig_2}
\end{figure}

\begin{figure}[ht]
\centering
\includegraphics[width=4.8in]{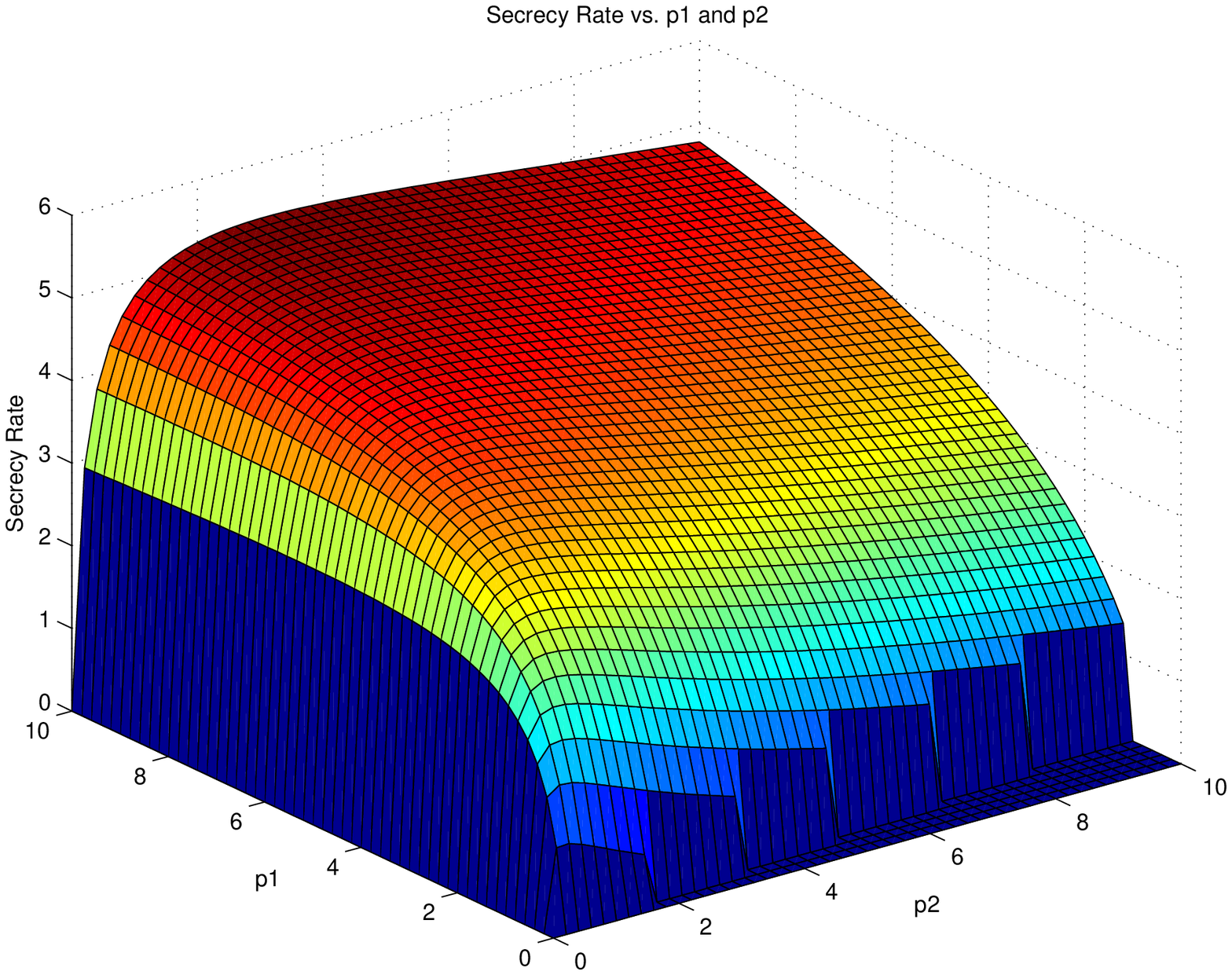}
\caption{Secrecy rate vs. $p_1$ and $p_2$  for the case without
jammers when ${g_{{S_1},R}} < {g_{{S_2},R}}$} \label{fig_3}
\end{figure}

In addition, we verify these optimal solutions by simulations in
different cases which is shown in Fig.~\ref{fig_2} and
Fig.~\ref{fig_3} in Section-V, and after further calculation based
on the simulation results, we can get that the results agree with
the optimal power solutions well.

\section{Physical Layer Security with Friendly Jammers}%

In this section, through further analysis, we first find that the
secrecy rate of the sources can be effectively improved by utilizing
proper jamming power from the friendly jammers. Then, the problem
comes to how to control the jamming power from different friendly
jammers when optimizing the secrecy rate of the sources. In general,
in a cooperative wireless network with selfish nodes, nodes may not
serve a common goal or belong to a single authority. Thus, a
mechanism of reimbursement to the friendly jammers should be
employed such that the friendly jammers can earn benefits from
spending their own transmitting power in helping the sources for
secure data transmission. For the source side, the sources aim to
achieve the best performance of secrecy rate with the friendly
jammers' help with the least reimbursements to them. For the
friendly jammer side, each friendly jammer aims to earn the payment
not only covers its transmitting cost but also gains as many extra
profits as possible. Therefore, we employ a Stackelberg game model
\cite{FT-GameTheory-1993} as a power control scheme jointly
considering both the benefits of the sources and the friendly
jammers. In the Stackelberg game model we proposed, the two sources
as a unity is the sole buyer that starts the process of the proposed
Stackelberg game, and the friendly jammers are the sellers,
therefore, the sources are treated as leader while the friendly
jammers are the followers. Furthermore, the optimal solutions of the
jamming power and asking price are investigated and a corresponding
distributed updating algorithm is provided. Finally, a centralized
scheme is proposed for performance comparison.

\subsection{Secrecy Rate Improvement using Friendly Jammers}

From (\ref{eq13}) and (\ref{eq14}), we have
\begin{align} \label{eq23}
C_1^s = \frac{W}{2}\left[ \log \left( {1 +
\frac{{{p_1}{g_{{S_1},R}}}}{{{\sigma ^2} + {K_1} + \sum\limits_i
{\frac{{{\sigma ^2}{g_{{J_i},R}}}}{{{p_r}{g_{{S_2},R}}}}p_i^J} }}}
\right) \right. \nonumber \\
\left. - \log \left( {1 + \frac{{{p_1}{g_{{S_1},R}}}}{{{\sigma ^2} +
{p_2}{g_{{S_2},R}} + \sum\limits_i {{g_{{J_i},R}}p_i^J} }}} \right)
\right]^ +,
\end{align}
and
\begin{align} \label{eq24}
C_2^s = \frac{W}{2}\left[ \log \left( {1 +
\frac{{{p_2}{g_{{S_2},R}}}}{{{\sigma ^2} + {K_2} + \sum\limits_i
{\frac{{{\sigma ^2}{g_{{J_i},R}}}}{{{p_r}{g_{{S_1},R}}}}p_i^J} }}}
\right) \right. \nonumber \\
\left. - \log \left( {1 + \frac{{{p_2}{g_{{S_2},R}}}}{{{\sigma ^2} +
{p_1}{g_{{S_1},R}} + \sum\limits_i {{g_{{J_i},R}}p_i^J} }}} \right)
\right]^ +.
\end{align}

From (\ref{eq23}) and (\ref{eq24}), we can see that both ${C_k}$ and
$C_k^m$, $k = 1,2$, are decreasing and convex functions of jamming
power $p_i^J$, $i\in\mathcal{N}$. However, if $C_k^m$ decreases
faster than ${C_k}$ as the jamming power $p_i^J$ increases, $C_k^s$
might increase in some region of value $p_i^J$. But when $p_i^J$
further increases, both ${C_k}$ and $C_k^m$ will approach zero. As a
result, $C_k^s$ approaches zero. Compared to (\ref{eq15}) and
(\ref{eq16}), we can get that if $\frac{{{\sigma
^2}{g_{{J_i},R}}}}{{{p_r}{g_{{S_2},R}}}} < {g_{{J_i},R}}$ and
$\frac{{{\sigma ^2}{g_{{J_i},R}}}}{{{p_r}{g_{{S_1},R}}}} <
{g_{{J_i},R}}$, $\forall i\in\mathcal{N}$, i.e., $\frac{{{\sigma
^2}}}{{{p_r}}} < \min \left\{ {{g_{{S_1},R}},{g_{{S_2},R}}}
\right\}$, the gain of the secrecy rate will be above zero in some
region of the jamming power $p_i^J$. Then the problem comes to how
to utilize the jamming power from different friendly jammers
effectively to maximize the secrecy rate. Thus, we propose a
Stackelberg game model to achieve effective jamming power control in
the following subsections.

Note that synchronization among the sources and the friendly jammers
is important in the investigated system with friendly jammers. Many
works have been devoted to the synchronization issues among
distributed nodes in cooperative networks, for example in
\cite{R1,R2}, effective synchronization schemes among distributed
sensors and cooperative relays with low complexity and good
performance were proposed. Thus, the synchronization issue among the
sources and the friendly jammers can be addressed effectively using
methods similar to those proposed in \cite{R1,R2}. However, this is
not the key investigated issue in this paper, therefore, we assume
that perfect synchronization among the sources and the friendly
jammers is implemented in the system.

\subsection{Source Side Game}

We consider the two sources as two buyers who want to optimize their
secrecy rates, while the cost paid for the ``service'', i.e.,
jamming power $p_i^J$, $i\in\mathcal{N}$, should also be taken into
consideration. For the source side we can define the utility
function as
\begin{align} \label{eq25}
{U_s} = a\left( {C_1^s + C_2^s} \right) - M,
\end{align}
where $a$ is a positive constant representing the economic gain per
unit rate of confidential data transmission between the two sources,
and $M$ is the cost to pay for the friendly jammers. Here we have
\begin{align} \label{eq26}
M = \sum\limits_{i\in\mathcal{N}} {{m_i}p_i^J},
\end{align}
where $m_i$ is the price per unit power paid for the friendly jammer
$i$ by the sources, $i\in\mathcal{N}$.


When considering the optimal transmitting power vector of source
$S_1$ and $S_2$, i.e., $\left( {{p_1},{p_2}} \right)$ to achieve the
maximum utility value in (\ref{eq25}), we can treat the jamming
power $p_i^J$, $i\in\mathcal{N}$, as constants since all the nodes
transmit with independent power. Thus, we can obtain similar results
of optimal power solutions as given in Subsection-III-B. But to
obtain the optimal solutions of $p_1$ and $p_2$ is not our main
purpose here. In this subsection, we formulate the source side game
to study how to effectively utilize the jamming power from different
friendly jammers in order to achieve the maximum utility value.

Then the source side game can be expressed as
\begin{align} \label{eq27}
&\max \;{U_s} = \max \;\left( {a\left( {C_1^s + C_2^s} \right) - M} \right), \\
\mbox{s.t.}\quad &\left\{ \begin{array}{l}
C_1^s \ge 0,C_2^s \ge 0 \\
0 \le p_i^J \le {p_{max}},{p_r} = {p_{max}},fixed\,{p_1},{p_2} \\
\end{array} \right.. \nonumber
\end{align}

The goal of the sources as buyers is to buy the optimal amount of
power from the friendly jammers in order to maximize the secrecy
rate. From (\ref{eq23}), (\ref{eq24}), and (\ref{eq27}), we have
\begin{align} \label{eq28}
{U_s} = &\frac{{aW}}{2}\left( {\log \frac{{1 + \frac{1}{{{A_1} +
\sum\limits_i {{T_{1,i}}p_i^J} }}}}{{1 + \frac{1}{{{B_1} +
\sum\limits_i {{L_{1,i}}p_i^J} }}}} + \log \frac{{1 +
\frac{1}{{{A_2} + \sum\limits_i {{T_{2,i}}p_i^J} }}}}{{1 +
\frac{1}{{{B_2} + \sum\limits_i {{L_{2,i}}p_i^J} }}}}} \right) -
\sum\limits_i {{m_i}} p_i^J,
\end{align}
where ${A_1} \triangleq \frac{{{\sigma ^2} +
{K_1}}}{{{p_1}{g_{{S_1},R}}}}$, ${A_2} \triangleq \frac{{{\sigma ^2}
+ {K_2}}}{{{p_2}{g_{{S_2},R}}}}$, ${B_1} \triangleq \frac{{{\sigma
^2} + {p_2}{g_{{S_2},R}}}}{{{p_1}{g_{{S_1},R}}}}$, ${B_2} \triangleq
\frac{{{\sigma ^2} + {p_1}{g_{{S_1},R}}}}{{{p_2}{g_{{S_2},R}}}}$,
${T_{1,i}} \triangleq \frac{{{\sigma
^2}{g_{{J_i},R}}}}{{{p_r}{p_1}{g_{{S_2},R}}{g_{{S_1},R}}}}$,
${T_{2,i}} \triangleq \frac{{{\sigma
^2}{g_{{J_i},R}}}}{{{p_r}{p_2}{g_{{S_2},R}}{g_{{S_1},R}}}}$,
${L_{1,i}} \triangleq \frac{{{g_{{J_i},R}}}}{{{p_1}{g_{{S_1},R}}}}$,
and ${L_{2,i}} \triangleq
\frac{{{g_{{J_i},R}}}}{{{p_2}{g_{{S_2},R}}}}$, $i\in\mathcal{N}$.

By differentiating (\ref{eq28}) with respect to $p_i^J$, we have
\begin{align} \label{eq29}
\frac{{\partial {U_s}}}{{\partial p_i^J}} =  &-
\frac{{aW{T_{1,i}}}}{{2\left( {{A_1} + \sum\limits_i
{{T_{1,i}}p_i^J} } \right)\left( {1 + {A_1} + \sum\limits_i
{{T_{1,i}}p_i^J} } \right)}} \nonumber \\
&+ \frac{{aW{L_{1,i}}}}{{2\left( {{B_1} + \sum\limits_i {{L_{1,i}}p_i^J} } \right)\left( {1 + {B_1} + \sum\limits_i {{L_{1,i}}p_i^J} } \right)}}  \nonumber \\
&- \frac{{aW{T_{2,i}}}}{{2\left( {{A_2} + \sum\limits_i
{{T_{2,i}}p_i^J} } \right)\left( {1 + {A_2} + \sum\limits_i
{{T_{2,i}}p_i^J} } \right)}} \nonumber \\
&+ \frac{{aW{L_{2,i}}}}{{2\left( {{B_2} + \sum\limits_i
{{L_{2,i}}p_i^J} } \right)\left( {1 + {B_2} + \sum\limits_i
{{L_{2,i}}p_i^J} } \right)}}{\kern 1pt}  - {m_i}.
\end{align}

Rearranging the above equation, when $\frac{{\partial
{U_s}}}{{\partial p_i^J}} = 0$, we can get an eighth order
polynomial equation as
\begin{align} \label{eq30}
{\left( {p_i^J} \right)^8} + {F_{i,7}}{\left( {p_i^J} \right)^7} +
{F_{i,6}}{\left( {p_i^J} \right)^6} + {F_{i,5}}{\left( {p_i^J}
\right)^5} + {F_{i,4}}{\left( {p_i^J} \right)^4} \nonumber \\ +
{F_{i,3}}{\left( {p_i^J} \right)^3} + {F_{i,2}}{\left( {p_i^J}
\right)^2} + {F_{i,1}}p_i^J + {F_{i,0}} = 0,
\end{align}
where ${F_{i,l}}$, $l=0,1,\ldots,7$, are formulae of constants
${A_k}$, ${B_k}$, ${T_{i,k}}$, ${L_{i,k}}$, and variables $p_j^J$,
$k=1,2$, $i\in\mathcal{N}$, $j\in\mathcal{N}$ but $j \ne i$.

The solutions to the high order equation (\ref{eq30}) can be
expressed in closed form, but the expressions of the solutions are
extremely complex and have little necessity for our following work.
Actually, what to our particular interest are not the closed-form
expressions of the optimal jamming power, but the parameters that
affect these optimal solutions. Thus, the optimal jamming power
solution can be expressed as
\begin{align} \label{eq31}
{p_i^J}^* = {p_i^J}^*\left( {{m_i},\left\{ {{A_k}} \right\},\left\{
{{B_k}} \right\},\left\{ {{T_{k,i}}} \right\},\left\{ {{L_{k,i}}}
\right\},{{\left\{ {p_j^J} \right\}}_{j \ne i}}} \right),
\end{align}
which is a function of the friendly jammer's price $m_i$, the other
jammers' jamming power ${\left\{ {p_j^J} \right\}_{j \ne i}}$, and
other system parameters. Noting that there may be up to eight roots
of the polynomial equation (\ref{eq30}), the selected solution
should be a real root and can lead to a higher value of $U_s$ in
(\ref{eq28}) than the other real ones. Subject to the power
constraints $0 \le p_i^J \le {p_{max}}$ in the game, we can get the
optimal strategy as
\begin{align} \label{eq32}
p_{i\_opt}^J = \min \left( {\max \left( {p{{_i^J}^*},0}
\right),{p_{max}}} \right).
\end{align}
If there are no real roots of the equation (\ref{eq30}), then the
optimal strategy will be either $p_{i\_opt}^J = 0$ or $p_{i\_opt}^J
= p_{max}$ according to which one can achieve a larger $U_s$ when
other parameters are settled.

Because of the high complexity of the solutions to the high order
equation in (\ref{eq31}), we further consider a special high
interference case to obtain a simple expression of the optimal
solution. In this special case, we assume that there is one jammer
staying very close to the malicious relay, so that the interference
from the jammer is much stronger than the power of the received
signals from the sources at the relay. Meanwhile, we also assume
that the received signal power is much higher than the additive
noise, i.e., high signal-to-noise ratio, which means ${\sigma ^2}
\ll {p_1}{g_{{S_1},R}} \ll p_1^J{g_{{J_1},R}}$ and ${\sigma ^2} \ll
{p_2}{g_{{S_2},R}} \ll p_1^J{g_{{J_1},R}}$. Then, we have
$\frac{\sigma^2}{{p_1}{g_{{S_1},R}}} \ll 1$,
$\frac{\sigma^2}{{p_2}{g_{{S_2},R}}} \ll 1$,
$\frac{\sigma^2}{p_1^J{g_{{J_1},R}}} \ll 1$,
$\frac{{p_1}{g_{{S_1},R}}}{p_1^J{g_{{J_1},R}}} \ll 1$, and
$\frac{{p_1}{g_{{S_1},R}}}{p_1^J{g_{{J_1},R}}} \ll 1$. We assume all
the left sides of these inequalities which are much smaller than $1$
approach zero. Therefore, the utility function of the source side in
(\ref{eq25}) can be approximately calculated as
\begin{align} \label{eq33}
{U_s} \approx &\frac{{aW}}{2}\left( {\log \left( {1 +
\frac{{{p_r}{p_1}{g_{{S_1},R}}{g_{{S_2},R}}}}{{{\sigma
^2}p_1^J{g_{{J_1},R}}}}} \right)} - \log \left( {1 +
\frac{{{p_1}{g_{{S_1},R}}}}{{p_1^J{g_{{J_1},R}}}}} \right) \right.
\nonumber \\
&\left. + \log \left( {1 + \frac{{{p_r}{p_2}{g_{{S_1},R}}{g_{{S_2},R}}}}{{{\sigma ^2}p_1^J{g_{{J_1},R}}}}} \right) - \log \left( {1 + \frac{{{p_2}{g_{{S_2},R}}}}{{p_1^J{g_{{J_1},R}}}}} \right) \right) - {m_1}p_1^J \nonumber \\
\approx &\frac{{aW}}{2}\left( {\frac{{{p_r}{p_1}{g_{{S_1},R}}{g_{{S_2},R}}}}{{{\sigma ^2}p_1^J{g_{{J_1},R}}}} - \frac{{{p_1}{g_{{S_1},R}}}}{{p_1^J{g_{{J_1},R}}}} + \frac{{{p_r}{p_2}{g_{{S_1},R}}{g_{{S_2},R}}}}{{{\sigma ^2}p_1^J{g_{{J_1},R}}}} - \frac{{{p_2}{g_{{S_2},R}}}}{{p_1^J{g_{{J_1},R}}}}} \right) - {m_1}p_1^J \nonumber \\
= &\frac{{aW}}{2}\left( {\left( {\frac{{{p_r}{g_{{S_2},R}}}}{{{\sigma ^2}}} - 1} \right)\frac{{{p_1}{g_{{S_1},R}}}}{{{g_{{J_1},R}}}} + \left( {\frac{{{p_r}{g_{{S_1},R}}}}{{{\sigma ^2}}} - 1} \right)\frac{{{p_2}{g_{{S_2},R}}}}{{{g_{{J_1},R}}}}} \right)\frac{1}{{p_1^J}} - {m_1}p_1^J \nonumber \\
= &- {D_1}\frac{1}{{p_1^J}} - {m_1}p_1^J,
\end{align}
where ${D_1} = \frac{{aW}}{2}\left[ {\left( {1 -
\frac{{{p_r}{g_{{S_2},R}}}}{{{\sigma ^2}}}}
\right)\frac{{{p_1}{g_{{S_1},R}}}}{{{g_{{J_1},R}}}} + \left( {1 -
\frac{{{p_r}{g_{{S_1},R}}}}{{{\sigma ^2}}}}
\right)\frac{{{p_2}{g_{{S_2},R}}}}{{{g_{{J_1},R}}}}} \right]$ and
the second approximation comes from the Taylor series expansion
$\log \left( {1 + x} \right) \approx x$ when $x$ is small enough
\footnote{Here we say $x$ is small enough means the high order of
$x$ approaches zero.}. It can be easily observed that if ${D_1} <
0$, $U_s$ is a decreasing function of $p_1^J$. As a result, $U_s$
can be optimized when $p_1^J = 0$, i.e., the jammer would not play
in the game. If ${D_1} > 0$, in order to find the optimal power for
the sources to buy, we can calculate
\begin{align} \label{eq34}
\frac{{\partial {U_s}}}{{\partial p_1^J}} = \frac{{{D_1}}}{{{{\left(
{p_1^J} \right)}^2}}} - {m_1} = 0.
\end{align}
Hence, the optimal closed-form solution can be expressed as
\begin{align} \label{eq35}
{p{_1^J}^*} = \sqrt {\frac{{{D_1}}}{{{m_1}}}}.
\end{align}

By comparing ${p{_1^J}^*}$ with the power under the boundary conditions, we can obtain the optimal solution of the source side game for this special case as
\begin{align} \label{eq36}
p_{1\_opt}^J = \min \left( {p{{_1^J}^*} , {p_{max}}} \right).
\end{align}

In Section-\uppercase\expandafter{\romannumeral5}, we employ the
general case setups for simulation. The results indicates that the
sources always prefer to buying power from only one jammer when
there exists at least one sufficiently-effective jammer, which is
more effective to perform jamming than the other jammers and is
simultaneously asking for a proper price. Therefore, this special
case with one jammer experiencing severe interference is valid in
analyzing the proposed game. Under this special case, we can get a
property of the proposed game that the optimal power consumption
${p{_1^J}^*}$ is a monotonous function of its price $m_1$, which
could help to prove the existence of the equilibrium in
Subsection-\uppercase\expandafter{\romannumeral4}-D. We can also
prove that the friendly jammer power ${p{_1^J}^*}$ bought from the
sources is convex in the price $m_1$ under some conditions.

From (\ref{eq35}), we have the first order derivative of
${p{_1^J}^*}$ as
\begin{align} \label{eq37}
\frac{{\partial p{{_1^J}^*}}}{{\partial {m_1}}} =  -
\frac{1}{2}\sqrt {{D_1}} m_1^{ - \frac{3}{2}},
\end{align}
and the second order derivative as
\begin{align} \label{eq38}
\frac{{{\partial ^2}p{{_1^J}^*}}}{{\partial {m_1}}} =
\frac{3}{4}\sqrt {{D_1}} m_1^{ - \frac{5}{2}} > 0,
\end{align}
which indicates in the high interference case, the optimal power ${p{_1^J}^*}$ is a convex function of the price $m_1$.

\subsection{Friendly Jammer Side Game}

For each friendly jammer, we can define the utility function as
\begin{align} \label{eq39}
{U_i} = {m_i}{\left( {p_i^J} \right)^{{c_i}}},
\end{align}
where ${c_i} \ge 1$ is a constant to balance the payment
${m_i}p_i^J$ from the sources and the transmission cost $p_i^J$ of
the jammer itself, $i\in\mathcal{N}$. With different values of
$c_i$, the jammers have different strategies for asking the price
$m_i$. Here, the jamming power $p_i^J$ is also a function of the
vector of prices $\left( {m_1,m_2,\ldots,m_N} \right)$, as the
amount of jamming power that the sources will buy also depends on
the prices that the friendly jammers ask. Hence, the friendly jammer
side game can be expressed as
\begin{align} \label{eq40}
\mathop {\max }\limits_{{m_i}} \;{U_i}.
\end{align}

The goal of each friendly jammer as a seller is to set an optimal
price in order to maximize its utility. By differentiating the
utility in (\ref{eq39}) and setting it to zero, we can get
\begin{align} \label{eq41}
\frac{{\partial {U_i}}}{{\partial {m_i}}} = {\left( {p_{i\_opt}^J}
\right)^{{c_i}}} + {m_i}{c_i}{\left( {p_{i\_opt}^J} \right)^{{c_i} -
1}}\frac{{\partial p_{i\_opt}^J}}{{\partial {m_i}}} = 0,
\end{align}
which is equivalent to solve
\begin{align} \label{eq42}
{\left( {p_{i\_opt}^J} \right)^{{c_i} - 1}}\left( {p_{i\_opt}^J +
{m_i}{c_i}\frac{{\partial p_{i\_opt}^J}}{{\partial {m_i}}}} \right)
= 0.
\end{align}
The equation (\ref{eq42}) can be solved by setting either
$p_{i\_opt}^J = 0$ or
\begin{align} \label{eq43}
p_{i\_opt}^J + {m_i}{c_i}\frac{{\partial p_{i\_opt}^J}}{{\partial
{m_i}}} = 0.
\end{align}
Hence, with the optimal solution $p_{i\_opt}^J$, we can get the
solution of the optimal price $m_{i\_opt}$ as a function given as
\begin{align} \label{eq44}
{m_{i\_opt}} = m_i^*\left\{ {{\sigma
^2},{g_{{S_1},R}},{g_{{S_2},R}},\left\{ {{g_{{J_i},R}}} \right\}}
\right\},
\end{align}
where ${m_{i\_opt}}$ should be positive; otherwise, the friendly
jammer would not participate the game, $i\in\mathcal{N}$.

\subsection{Stackelberg Equilibrium of the Proposed Game}

In this subsection, we investigate the Stackelberg equilibrium of
the proposed game, at which neither the sources nor each friendly
jammer can further improve its utility by changing its own strategy
only. From the game definitions of the source side in (\ref{eq27})
and the friendly jammer side in (\ref{eq40}), we can define the
Stackelberg equilibrium as follows:

\emph{Definition 1:} ${p_{i\_SE}^J}$ and $m_{i\_SE}$ are the
Stackelberg equilibrium of the proposed game, if when $m_i$ is
fixed,
\begin{align} \label{eq45}
{U_s}\left( {\left\{ {p_{i\_SE}^J} \right\}} \right) = \mathop {\sup
}\limits_{0 \le \left\{ {p_i^J} \right\} \le {p_{max}},\forall i}
{U_s}\left( {\left\{ {p_i^J} \right\}} \right), \forall
i\in\mathcal{N},
\end{align}
and when $p_i^J$ is fixed,
\begin{align} \label{eq46}
{U_i}\left( {{m_{i\_SE}}} \right) = \mathop {\sup }\limits_{{m_i}}
\; {U_i}\left( {{m_i}} \right), \forall i\in\mathcal{N}.
\end{align}

From the analysis in the previous two subsections, we have that
$\left\{ {p_{i\_opt}^J} \right\}_{i = 1}^N$ in (\ref{eq32}) and
$\left\{ {m_{i\_opt}} \right\}_{i = 1}^N$ in (\ref{eq44}) are the
optimal solutions of the jamming power needed by the sources and the
asking prices given by the friendly jammers when solving the utility
optimization problem in (\ref{eq27}) and (\ref{eq40}). And thus, we
can obtain the property that the pair of $\left\{ {p_{i\_opt}^J}
\right\}_{i = 1}^N$ and $\left\{ {m_{i\_opt}} \right\}_{i = 1}^N$ is
the Stackelberg equilibrium of the proposed game.

We can easily prove this property theoretically in the special high
interference case with one efficient friendly jammer close to the
untrusted relay. In Subsection-IV-B, we have proved that in this
special case, the optimal jamming power solution $p_{1\_opt}^J$
bought from the efficient friendly jammer $1$ is monotone decreasing
and convex with the asking price $m_1$, when the other friendly
jammers' prices are fixed. And in this case, the sources would
prefer to buy the jamming power only from the efficient friendly
jammer. Therefore, we can obtain that there exists a unique
Stackelberg equilibrium that are just the optimal solutions of the
jamming power and the asking prices. However, due to the extremely
complex closed-form expressions of the optimal solutions, for more
general cases, the proof in theory is intractable, and thus instead,
we prove by simulations in Section-V that the proposed game can
effectively converge to a unique Stackelberg equilibrium, which are
the optimal solutions of the jamming power and the asking price when
maximizing the utilities of the sources and the friendly jammers.

\subsection{Distributed Updating Algorithm}

In this subsection, we construct a corresponding distributed
updating algorithm for the proposed game to converge to the
Stackelberg equilibrium defined above. By rearranging (\ref{eq43}),
we have
\begin{align} \label{eq47}
{m_i} = {I_i}\left( \textbf{m} \right) =  - \frac{{\left(
{p_{i\_opt}^J} \right)}}{{{c_i}\frac{{\partial
p_{i\_opt}^J}}{{\partial {m_i}}}}},
\end{align}
where $\textbf{m} = {\left[ {{m_1},{m_2},\ldots,{m_N}} \right]^T}$,
${p_{i\_opt}^J} $ is a function of $\textbf{m}$, and ${I_i}\left(
\textbf{m} \right)$ is the price update function for friendly jammer
$i$, $i\in\mathcal{N}$. The information for updating can be obtained
from the sources, which is similar to the distributed power
allocation \cite{HL-ResourceAllocation-2008}. The distributed
algorithm can be expressed in a vector form as
\begin{align} \label{eq48}
\textbf{m}\left( {t + 1} \right) = \textbf{I}\left(
{\textbf{m}\left( t \right)} \right),
\end{align}
where $\textbf{I} = {\left[ {{I_1},{I_2},...,{I_N}} \right]^T}$, and
the iteration is from time $t$ to time $t+1$.

Furthermore, we can get the convergence of the proposed scheme by
proving that the update function in (\ref{eq48}) is a standard
function \cite{Yates-1995} defined as

\emph{Definition 2:} A function $\textbf{I}\left( \textbf{m}
\right)$ is standard, if for all $\textbf{m} \ge \textbf{0}$, the
following properties are satisfied

\begin{enumerate}

\item Positivity: $\textbf{I}\left( \textbf{m} \right) \ge \textbf{0}$,

\item Monotonicity: if $\textbf{m} \ge \textbf{m}'$, then $\textbf{I}\left( \textbf{m} \right) \ge \textbf{I}\left( {\textbf{m}'}
\right)$, or $\textbf{I}\left( \textbf{m} \right) \le
\textbf{I}\left( {\textbf{m}'} \right)$,

\item Scalability: For all $\eta > 1$, $\eta \textbf{I}\left( \textbf{m} \right) \ge \textbf{I}\left( {\eta \textbf{m}}
\right)$.

\end{enumerate}

Similarly as the situation in \cite{Yates-1995}, we can get that
each friendly jammer's price will converge to a fixed point, i.e.,
the Stackelberg equilibrium in our game, from any feasible initial
price vector $\textbf{m}\left( 0 \right)$. The positivity of the
update function is easy to prove. We know that if the price $m_i$
goes up, the sources would buy less from the friendly jammer $i$.
Therefore, ${\frac{{\partial p_{i\_opt}^J}}{{\partial {m_i}}}}$ in
(\ref{eq47}) is negative, and we can prove the positivity of the
price update function as ${I_i}\left( \textbf{m} \right) =  -
\frac{{\left( {p_{i\_opt}^J} \right)}}{{{c_i}\frac{{\partial
p_{i\_opt}^J}}{{\partial {m_i}}}}} > 0$, $\forall i\in\mathcal{N}$.

Because of the complexity of the optimal solutions in (\ref{eq32})
and (\ref{eq44}), the monotonicity and scalability can only be shown
in the high interference case. From (\ref{eq35}) and (\ref{eq47}),
we have the update function as
\begin{align} \label{eq49}
{I_1}\left( \textbf{m} \right) =  - \frac{{\left( {p{{_1^J}^*}}
\right)}}{{{c_1}\frac{{\partial p{{_1^J}^*}}}{{\partial {m_1}}}}} =
- \frac{{\sqrt {\frac{{{D_1}}}{{{m_1}}}} }}{{ -
\frac{{{c_1}}}{2}\sqrt {{D_1}} m_1^{ - \frac{3}{2}}}} =
\frac{{2{m_1}}}{{{c_1}}},
\end{align}
which is monotonically increasing with the price $m_1$ and scalable.

For more general cases, the analysis is intractable. But in
Section-\uppercase\expandafter{\romannumeral5}, we employ the
general case setups for simulation and the results show that the
proposed distributed game scheme can converge and outperform the
case without jammers.

\subsection{Centralized Scheme for Comparison}

Traditionally, the centralized scheme is employed assuming that all
the channel information is known. In this subsection, we formulate
the centralized problem by optimizing the secrecy rate with respect
to the constraints of maximal jamming power $p_{max}$.
\begin{align} \label{eq50}
&\mathop {\max }\limits_{p_i^J} \;{C^s} = \mathop {\max
}\limits_{p_i^J}\; \left( {C_1^s + C_2^s} \right), \\
\mbox{s.t.}\quad &0 \leq p_i^J \leq p_{max}, \forall i \in
\mathcal{N},  \nonumber
\end{align}
where $C_1^s$ and $C_2^s$ are obtained from (\ref{eq23}) and
(\ref{eq24}), respectively. The centralized solution is found by
maximizing the secrecy rate only.

In Section-\uppercase\expandafter{\romannumeral5}, we compare the
proposed distributed algorithm with this centralized scheme. From
the simulation results, we can see that the distributed solution and
the centralized solution are asymptotically the same when the unit
rate's gain $a$ in (\ref{eq27}) is sufficiently large. However, the
distributed solution only needs to update the difference of the
power and price to be adaptive, while the centralized solution
requires all channel information in each time slot. Therefore, the
distributed algorithm is more efficient in practical applications.

\section{Simulation Results}%

To investigate the performances, we conduct the following
simulations. For simplicity and without loss of generality, we
consider a two-way relay system where the sources $S_1$, $S_2$, and
the malicious relay $R$ are located at the coordinate $(-1,0)$,
$(1,0)$, and $(0,0)$, respectively. The other simulation parameters
are set up as follows: The maximum power constraint ${p_{max}}$ is
$10$; the transmission bandwidth is $1$; the noise variance is
${\sigma ^2} = 0.01$; Rayleigh fading channel is assumed, where the
channel gain consists of the path loss and the Rayleigh fading
coefficient; the path loss factor is $2$. Here, we select $a = 1$
for the source side utility in (\ref{eq27}).

For the special case without jammers, we set the jamming power up to
zero. In Fig.~\ref{fig_2} and Fig.~\ref{fig_3}, we show the secrecy
rate as a function of the two sources' transmitting power $p_1$ and
$p_2$ in this special case. It shows that the optimal power vector
of $\left( {{p_1},{p_2}} \right)$ is $\left(
{0.22{p_{max}},{p_{max}}} \right)$ when ${g_{{S_1},R}} = 0.3857$ and
${g_{{S_2},R}} = 0.0443$, and $\left( {{p_{max}},0.32{p_{max}}}
\right)$ when ${g_{{S_1},R}} = 0.0508$ and ${g_{{S_2},R}} = 0.3018$.
After further calculation, we can see that the results agree with
the optimal power allocation conclusions given in Section
\uppercase\expandafter{\romannumeral3}.

\begin{figure}[ht]
\centering
\includegraphics[width=4.8in]{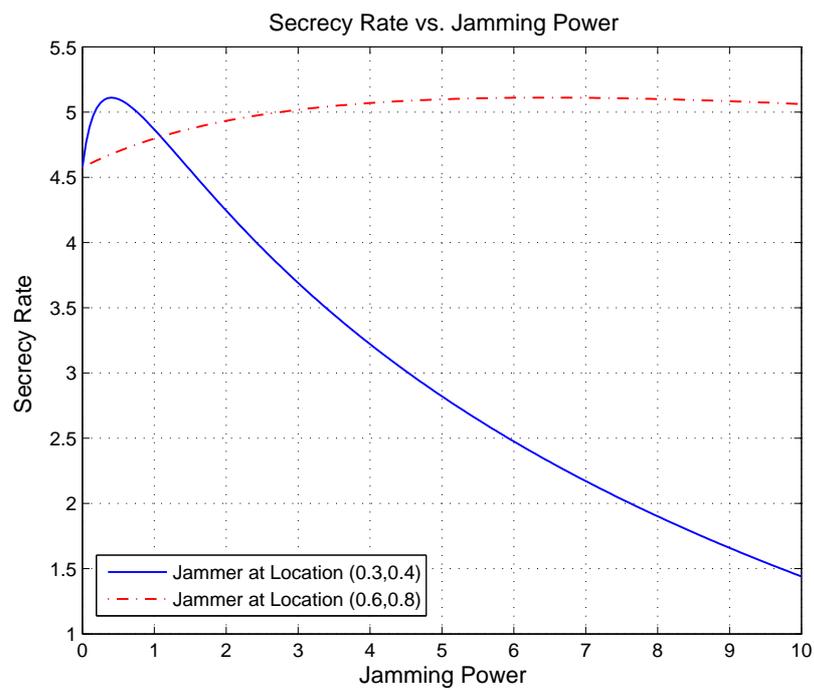}
\caption{Secrecy rate vs. jamming power} \label{fig_4}
\end{figure}

\begin{figure}[ht]
\centering
\includegraphics[width=4.8in]{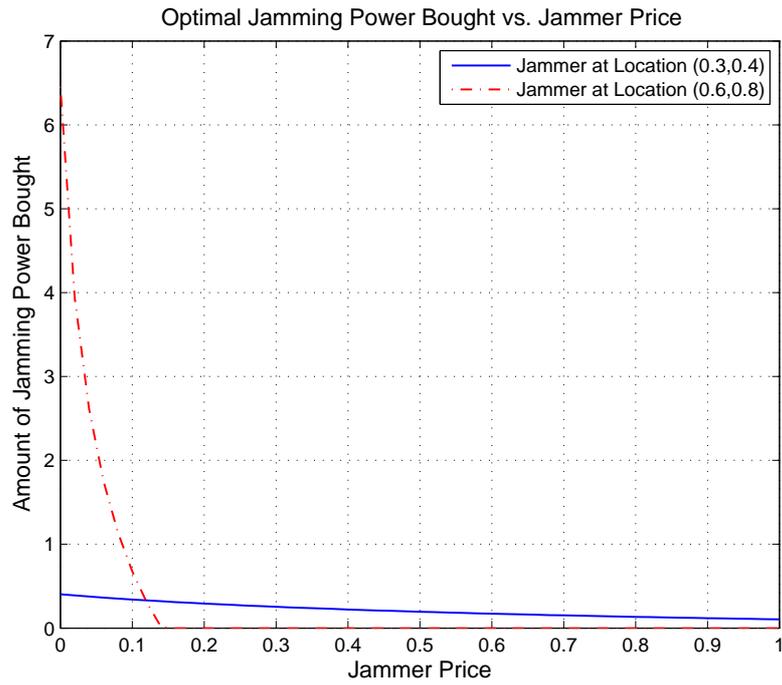}
\caption{How much jamming power the sources buy as a function of the
jammer price} \label{fig_5}
\end{figure}

For the single-jammer case, we consider two jammer locations which
are $(0.3,0.4)$ and $(0.6,0.8)$. Fig.~\ref{fig_4} shows the secrecy
rate as a function of the jamming power when $p_1$, $p_2$, and $p_r$
are all set up to ${p_{max}}$. We can see that with the increase of
the jamming power, the secrecy rate first increases and then
decreases. There indeed exists an optimal point for the jamming
power. Also the optimal point depends on the location of the
friendly jammer, and we can find that the friendly jammer close to
the malicious relay is more effective to improve the secrecy rate.
Fig.~\ref{fig_5} shows that the optimal amount of the jamming power
bought by the sources depends on the price requested by the friendly
jammer. We can see that the amount of bought power gets reduced if
the price goes high and the sources would even stop buying after
some price point. And thus there is a tradeoff for the jammers to
set the price. If the price is set too high, the sources would buy
less power or even stop buying.

\begin{figure}[ht]
\centering
\includegraphics[width=4.8in]{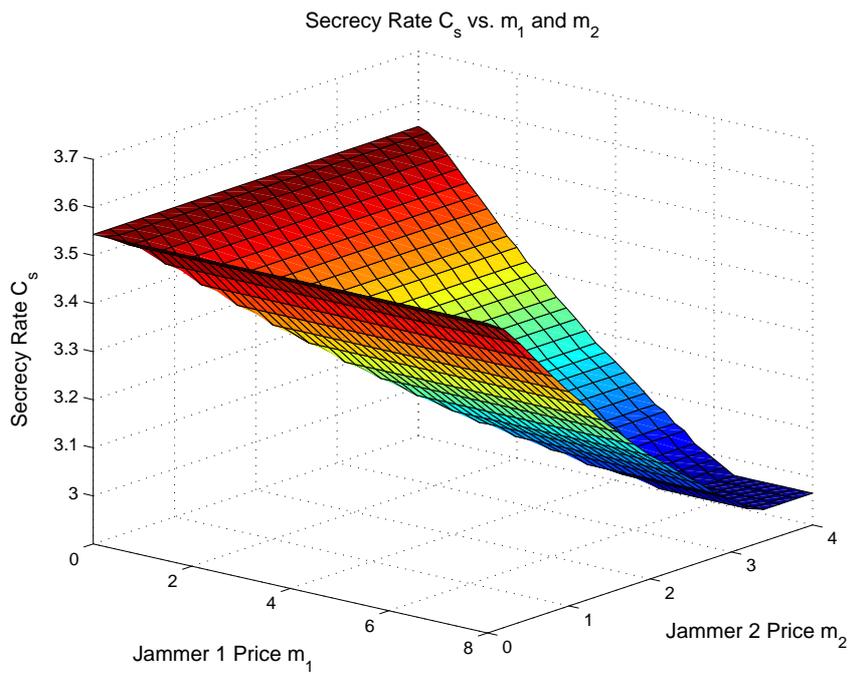}
\caption{Secrecy rate $C_s$ vs. the prices of both jammers}
\label{fig_6}
\end{figure}

\begin{figure}[ht]
\centering
\includegraphics[width=4.8in]{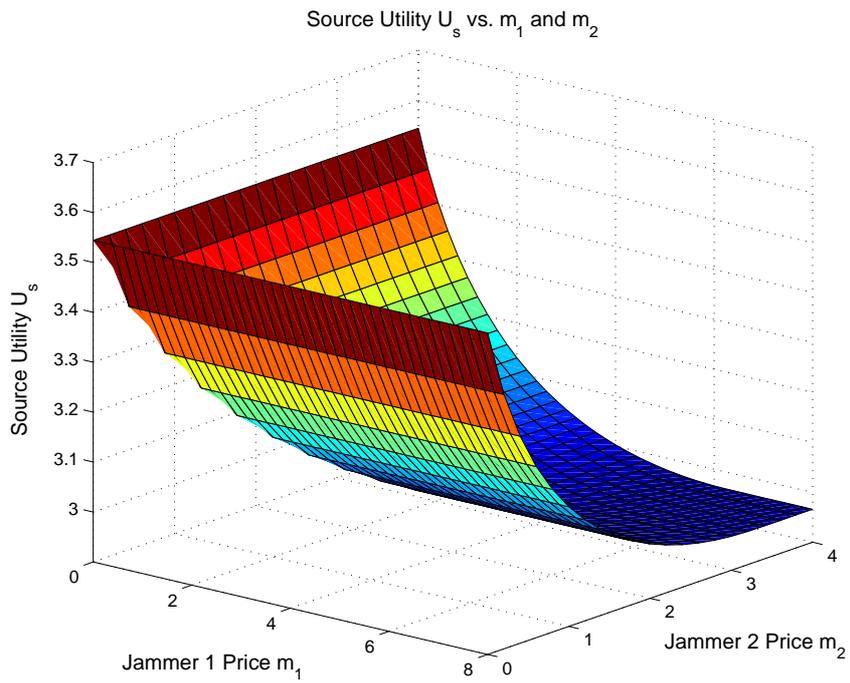}
\caption{The sources' utility $U_s$ vs. the prices of both jammers}
\label{fig_7}
\end{figure}

\begin{figure}[ht]
\centering
\includegraphics[width=4.8in]{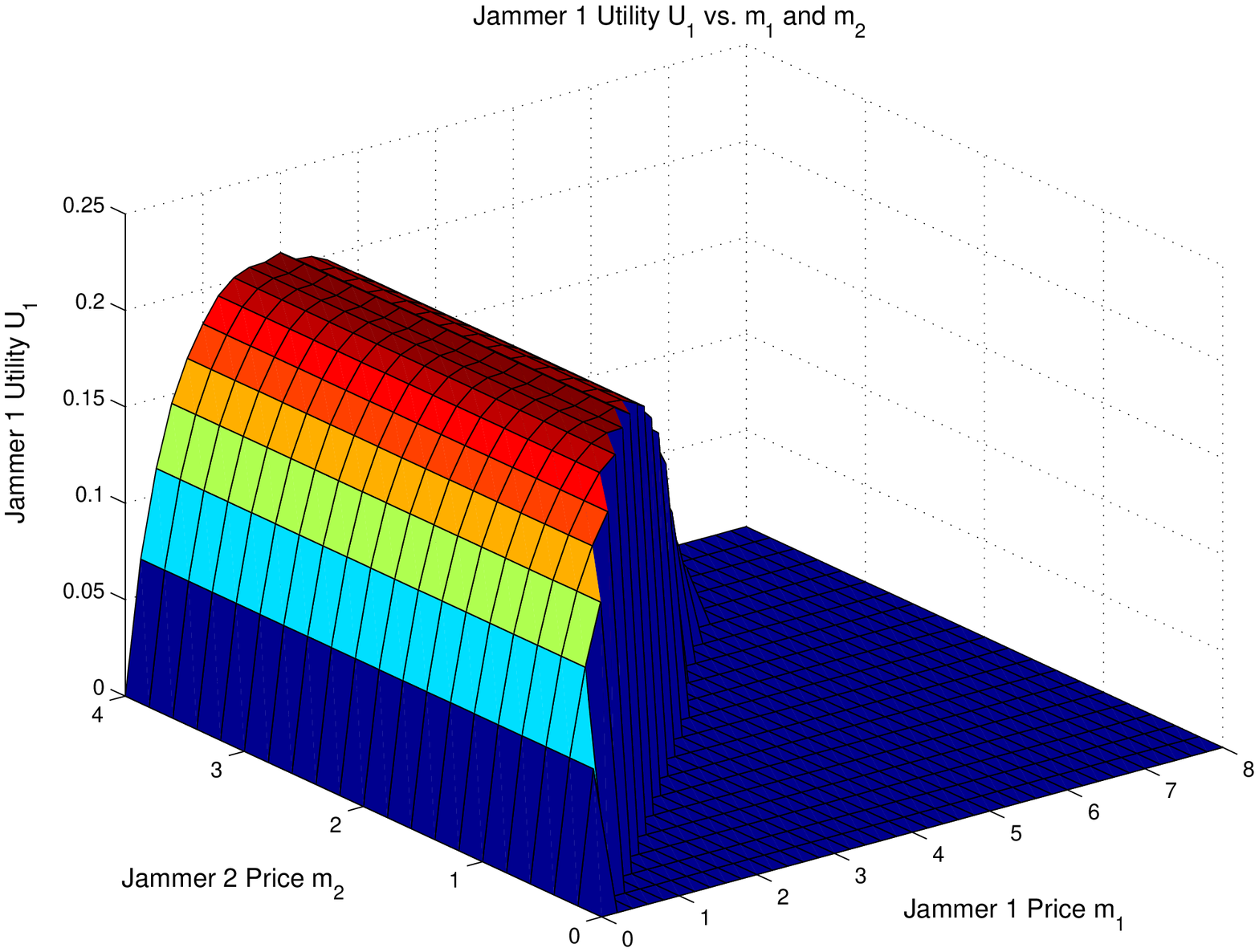}
\caption{The jammer 1's utility $U_1$ vs. the prices of both
jammers} \label{fig_8}
\end{figure}

\begin{figure}[ht]
\centering
\includegraphics[width=4.8in]{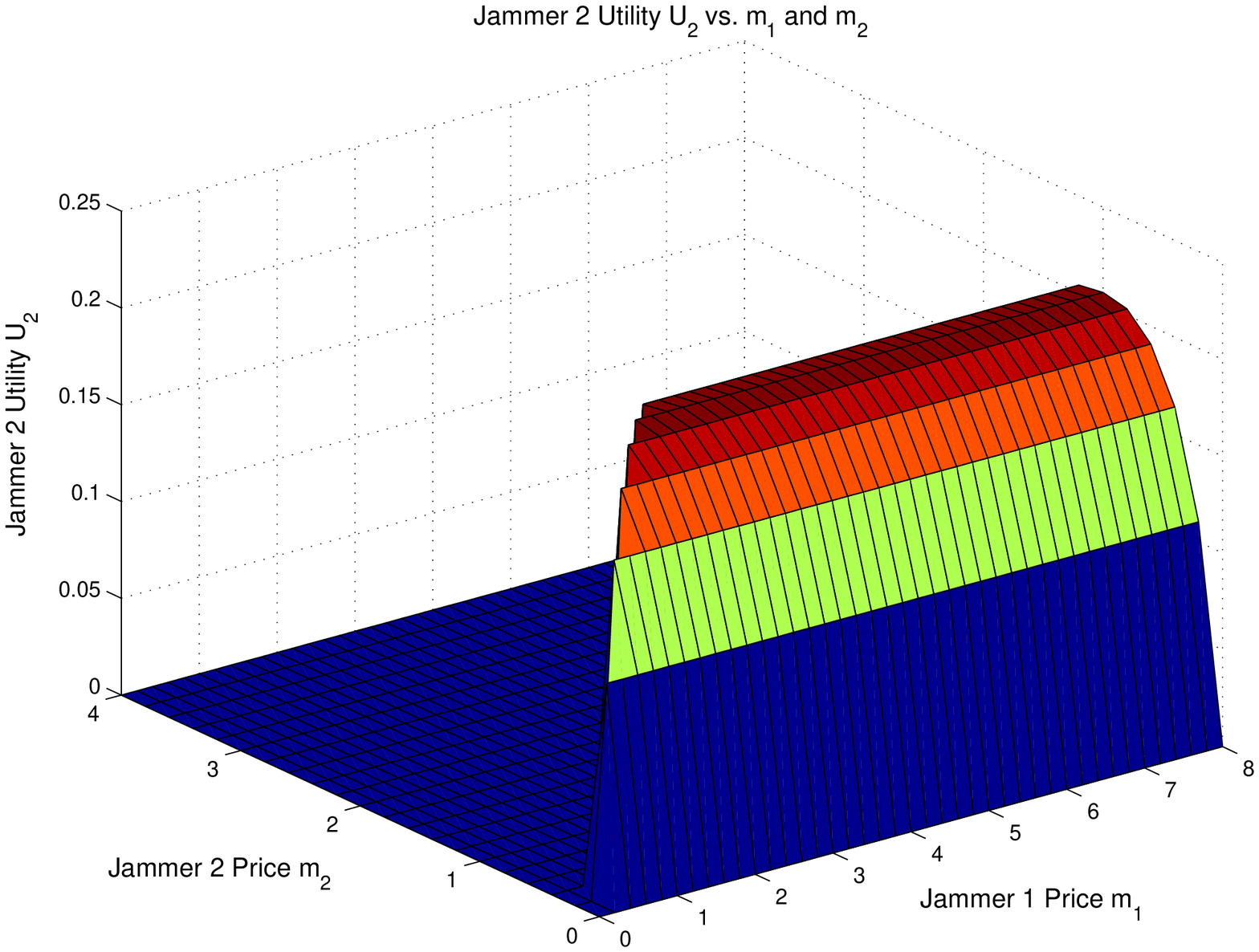}
\caption{The jammer 2's utility $U_2$ vs. the prices of both
jammers} \label{fig_9}
\end{figure}

For the multiple-jammer case, we consider two jammers which are
located at $(0.3,0.4)$ and $(0.5,0.5)$, respectively. Note that in
the simulations of this scenario the jammers are both
sufficiently-effective ones. Here we define jammer $i$ as a
sufficiently-effective one if it on its own can offer a power
$p_i^J$, $p_i^J \in \left( {0,{p_{max}}} \right]$, making the
secrecy rate improved up to the maximal value. In Fig.~\ref{fig_6},
Fig.~\ref{fig_7}, Fig.~\ref{fig_8}, and Fig.~\ref{fig_9}, the
secrecy rate $C_s$, the sources' utility $U_s$, the first jammer's
utility $U_1$, and the second jammer's utility $U_2$ as functions of
both jammers' prices are shown, respectively. In Fig.~\ref{fig_6},
we can see that there exists an upper bound and a lower bound for
the secrecy rate when the channel conditions are settled. When one
of the two jammers' prices is low enough, the sources could buy
sufficient jamming power from the jammer to improve the secrecy rate
up to the upper bound. When both jammers' prices are beyond the
sources' payment ability, the secrecy rate would stay at the lower
bound which is the same as the case without jammers as the sources
no longer buy jamming power from the jammers. In Fig.~\ref{fig_7},
we can see that if at least one of the two jammers sets a low price,
the sources' utility is high as the sources could get a high secrecy
rate at a low cost from the jammer with low asking price. With both
jammers' prices going high, the sources' utility decreases. When the
prices of both jammers are so high that the sources cannot benefit
any more from the jamming service at the high cost, the sources
would stop buying. In Fig.~\ref{fig_8} and Fig.~\ref{fig_9}, we can
find that under this condition (i.e., both of the jammers are
sufficiently-effective ones) the sources would always prefer to
buying service from only one of the friendly jammers for the best
performance. The selected jammer is either more effective to improve
the secrecy rate when the prices of the jammers are comparable, or
the one whose asking price is low enough. For the friendly jammer
side, if the jammer asks too low price, the jammer's utility is very
low. But if the jammer asks too high price, the sources might buy
the service from the other friendly jammer. Therefore, there is an
optimal price for each friendly jammer to ask, and the sources would
always select the one which can provide the best performance
improvement.

\begin{figure}[ht]
\centering
\includegraphics[width=4.8in]{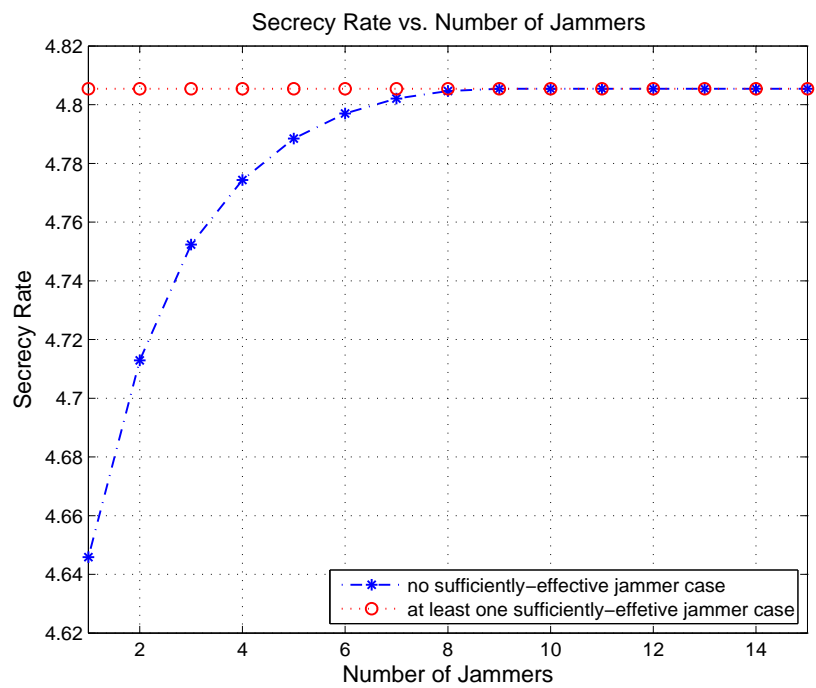}
\caption{Secrecy rate vs. number of jammers} \label{fig_10}
\end{figure}

For the multiple-jammer case, we also discuss how the optimal
secrecy rate changes with the number of jammers increasing. We
conduct simulations under two different scenarios, i.e., there
exists at least one sufficiently-effective jammer and there is no
sufficiently-effective jammer. Here no sufficiently-effective jammer
means that the sources could not achieve the maximal secrecy rate
with only one jammer's help. Therefore, the sources have to first
sort the current friendly jammers in an order of effectiveness, and
then buy jamming power from the most effective jammer one by one
until the secrecy rate reaches the maximal value. From (\ref{eq23})
and (\ref{eq24}), we can get that if the channel information and the
transmitting power of the sources and the relay are settled, the
maximal achievable secrecy rate will not change, no matter how many
jammers are used and how much jamming power bought by the sources
from each jammer. For the non-sufficiently-effective jammer
scenario, we set the jammer $i$ located at $(x_i,y_i)$,
$i\in\mathcal{N}$, where $x_i^2+y_i^2>4$. In Fig.~\ref{fig_10}, the
optimal secrecy rate as a function of number of friendly jammers is
shown. We can see that if there exists sufficiently-effective
jammers, the optimal secrecy rate does not change as the number of
jammers increases. For the sources, they always choose the most
effective jammer to achieve the maximal secrecy rate, when
additional jamming power from other jammers would decrease the
optimal secrecy rate. In other words, in this scenario, the optimal
secrecy rate can reach the maximal value with the most effective
jammer's help only. On the other hand, if there are no
sufficiently-effective jammers, the optimal secrecy rate will be
improved up to the maximal value as the number of jammers increases.

\begin{figure}[ht]
\centering
\includegraphics[width=4.8in]{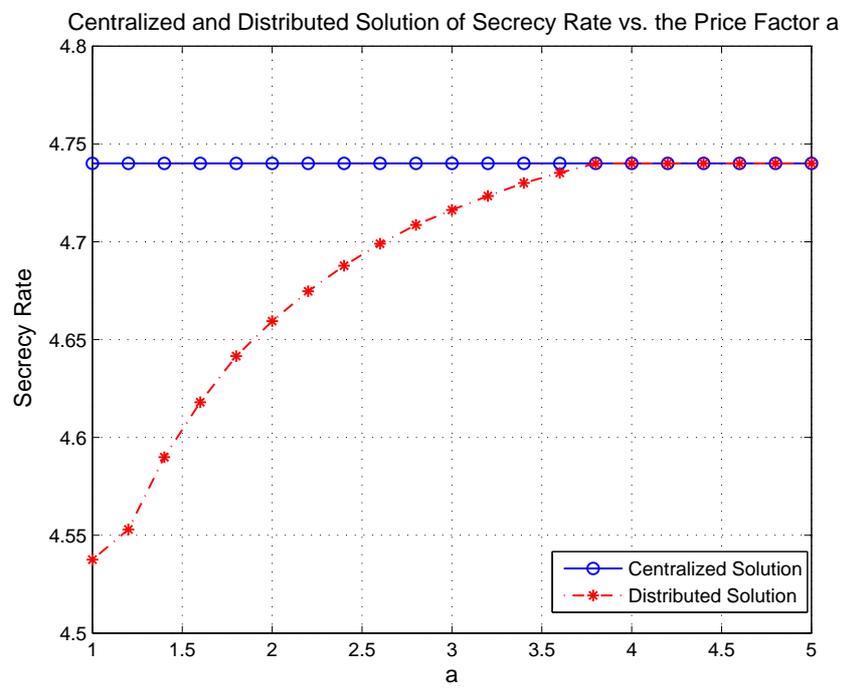}
\caption{Centralized and distributed solution of secrecy rate vs.
the gain factor} \label{fig_11}
\end{figure}

Finally, we compare the distributed solution with the centralized
solution of secrecy rate. In Fig.~\ref{fig_11}, we show the optimal
secrecy rate of both the distributed and centralized schemes as a
function of the gain factor $a$ in (\ref{eq25}) with the friendly
jammer located at $(1,1)$. When $a$ is large, the sources would care
the gain of secrecy rate more than the jamming service cost in the
game theoretical scheme. We can see that the performance gap between
the distributed and centralized solutions of optimal secrecy rate is
shrinking as $a$ is increasing. And the solutions are asymptotically
the same when $a$ is sufficiently large.

\section{Conclusions}%

In this paper, we have investigated the physical layer security for
two-way relay communications with untrusted relay and friendly
jammers. As a simple case, a two-way relay system without jammers is
first studied, and an optimal power allocation vector of the sources
and relay nodes is found. We then investigated the secrecy rate in
the presence of friendly jammers. Furthermore, we defined and
analyzed a Stackelberg type of game between the sources and the
friendly jammers to achieve the optimal secrecy rate in a
distributed way. Finally, we obtained a distributed solution for the
proposed game. From the simulation results we can get the following
conclusions. First, a non-zero secrecy rate for two-way relay
channel is indeed available. Second, the secrecy rate can be
improved with the help of friendly jammers, and there is an optimal
solution of jamming power allocation. Third, there is also a
tradeoff for the price a jammer sets, and if the price is too high,
the sources will turn to buying from others. For the game, we can
see that the distributed algorithm and the centralized scheme have
similar performances, especially when the gain factor is
sufficiently large.

\newpage

\end{document}